\documentclass[twocolumn,showpacs,aps,prd,floatfix]{revtex4}

\usepackage{graphicx}
\usepackage{dcolumn}
\usepackage{amsmath}
\usepackage{epsfig}
\usepackage{bm}
\usepackage{color}

\RequirePackage{xspace}

\usepackage{relsize}

\def\epem       {\ensuremath{e^+e^-}\xspace}
\def\pep2{PEP-II}
\def\KS    {\ensuremath{K^0_{\scriptscriptstyle S}}\xspace}
\def\Kbar  {\kern 0.2em\overline{\kern -0.2em K}{}\xspace}

\def\Kzb   {\ensuremath{\Kbar^0}\xspace}
\newcommand{\gev}{\ensuremath{\mathrm{\,Ge\kern -0.1em V}}\xspace}
\newcommand{\mev}{\ensuremath{\mathrm{\,Me\kern -0.1em V}}\xspace}
\newcommand{\gevc}{\ensuremath{{\mathrm{\,Ge\kern -0.1em V\!/}c}}\xspace}
\newcommand{\mevc}{\ensuremath{{\mathrm{\,Me\kern -0.1em V\!/}c}}\xspace}
\newcommand{\gevcc}{\ensuremath{{\mathrm{\,Ge\kern -0.1em V\!/}c^2}}\xspace}
\newcommand{\mevcc}{\ensuremath{{\mathrm{\,Me\kern -0.1em V\!/}c^2}}\xspace}

\def\invfb   {\ensuremath{\mbox{\,fb}^{-1}}\xspace}

\newcommand{\dedx}{\ensuremath{\mathrm{d}\hspace{-0.1em}E/\mathrm{d}x}\xspace}

\def\jetset74   {\mbox{\tt Jetset \hspace{-0.5em}7.\hspace{-0.2em}4}\xspace}

\def\p{\phantom{0}}

\def\babar{\mbox{\slshape B\kern-0.1em{\smaller A}\kern-0.1em B\kern-0.1em{\smaller A\kern-0.2em R}}\xspace}

\newcommand{\BABARPubYear}    {07}
\newcommand{\BABARPubNumber}  {042}
\newcommand{\SLACPubNumber} {12935}

\def\figurebox#1#2#3{%
    \def\arg{#3}%
    \ifx\arg\empty
    {\hfill\vbox{\hsize#2\hrule\hbox to #2{\vrule\hfill\vbox to #1{\hsize#2\vfill}\vrule}\hrule}\hfill}%
    \else
    {\hfill\epsfbox{#3}\hfill}%
    \fi}

\begin{document}

\begin{flushleft}
\babar-PUB-\BABARPubYear/\BABARPubNumber\\
SLAC-PUB-\SLACPubNumber\\
\end{flushleft}

\title{{\large \boldmath A Study of Excited Charm-Strange Baryons with Evidence for new Baryons $\Xi_c(3055)^+$ and $\Xi_c(3123)^+$}}

\author{B.~Aubert}
\author{M.~Bona}
\author{D.~Boutigny}
\author{Y.~Karyotakis}
\author{J.~P.~Lees}
\author{V.~Poireau}
\author{X.~Prudent}
\author{V.~Tisserand}
\author{A.~Zghiche}
\affiliation{Laboratoire de Physique des Particules, IN2P3/CNRS et Universit\'e de Savoie, F-74941 Annecy-Le-Vieux, France }
\author{J.~Garra~Tico}
\author{E.~Grauges}
\affiliation{Universitat de Barcelona, Facultat de Fisica, Departament ECM, E-08028 Barcelona, Spain }
% ...
\collaboration{The \babar\ Collaboration}
\noaffiliation

\begin{abstract}
We present a study of excited charm-strange baryon states produced in \epem annihilations at or near a center-of-mass energy of 10.58\gev, in a data sample with an integrated luminosity of 384\invfb recorded with the \babar detector at the \pep2 \epem storage rings at the Stanford Linear Accelerator Center.
We study strong decays of charm-strange baryons to
$\Lambda_c^+\KS$, $\Lambda_c^+K^-$, $\Lambda_c^+K^-\pi^+$, $\Lambda_c^+\KS\pi^-$, $\Lambda_c^+\KS\pi^-\pi^+$, and $\Lambda_c^+K^-\pi^+\pi^-$. 
This study confirms the existence of the states $\Xi_c(2980)^+$, $\Xi_c(3077)^+$, and $\Xi_c(3077)^0$, with a more accurate determination of the $\Xi_c(2980)^+$ mass and width.
We also present evidence for two new states, $\Xi_c(3055)^+$ and $\Xi_c(3123)^+$, decaying through the intermediate resonant modes $\Sigma_c(2455)^{++}K^-$ and $\Sigma_c(2520)^{++}K^-$, respectively.
For each of these baryons, we measure the yield in each final state, determine the statistical significance, and calculate the product of the production cross-section and branching fractions. 
We also measure the masses and widths of these excited charm-strange baryons.
\end{abstract}


\maketitle

\setcounter{footnote}{0}

\section{Introduction}
\label{sec:Introduction}
With the observation of the $\Omega_c^*$ ($css$) baryon~\cite{omega} by the \babar\ Collaboration, every predicted SU(4) ground-state single-charm baryon has been experimentally observed.
Several excited $\Lambda_c$ ($cqq$), $\Sigma_c$ ($cqq$), and $\Xi_c$ ($csq$) baryons have also been experimentally observed~\cite{PDG}.
The spins and parities of these are assigned based on a comparison of the measured masses and natural widths with predictions of theoretical models.

Both the \babar\ and Belle collaborations have searched for ground-state double-charm baryons decaying to the final state $\Lambda_c^+K^-\pi^+$~\cite{conj,dcb,belle2}.
These searches reveal no evidence for such states.
However, the Belle Collaboration finds evidence for two excited charm-strange baryon states, $\Xi_c(2980)^{+,0}$ and $\Xi_c(3077)^{+,0}$, decaying strongly to $\Lambda_c^+\KS\pi^-$ and $\Lambda_c^+K^-\pi^+$~\cite{belle2}.
Although these new states have the same or similar decay modes as those used in the search for weak decays of double-charm baryons, they are identified as charm-strange states based on the measured masses, natural widths (which indicate strong decays), and charges of the members of the isospin doublet. 

Previously known excited $\Xi_c$ baryons have been observed only in decays to a lower-mass $\Xi_c$ baryon plus a pion or photon.
In contrast, the $\Xi_c(2980)^{+,0}$ and $\Xi_c(3077)^{+,0}$ baryons are observed in decays in which the charm and strange quarks are in separate hadrons.
The observed decay modes may have implications for the internal quark dynamics of these new states.
Several excited charm-strange baryons with $J^P=\{(1/2)^{\pm},(3/2)^{\pm}\}$ are predicted, with masses ranging from about 2800\mevcc to 3150\mevcc~\cite{thspec,thlow}.
Other authors~\cite{post1,post2,post3} consider $J^P=(5/2)^+$ states and radial excitations, and use the measured natural widths and decay modes in assigning possible quantum numbers for the new $\Xi_c(2980)^{+,0}$ and $\Xi_c(3077)^{+,0}$ states. 

In this paper, we report measurements of excited charm-strange baryon properties based on a data sample corresponding to an integrated luminosity of 384\invfb recorded with the \babar detector at the \pep2 asymmetric-energy \epem storage rings at the Stanford Linear Accelerator Center.
We search for decays to the three-body final states $\Lambda_c^+K^-\pi^+$ and $\Lambda_c^+\KS\pi^-$,
the two-body final states $\Lambda_c^+\KS$ and $\Lambda_c^+K^-$,
and the four-body final states $\Lambda_c^+\KS\pi^-\pi^+$ and $\Lambda_c^+K^-\pi^-\pi^+$.
Significant signals are found only in decays to three-body final states.
We confirm the existence of the states $\Xi_c(2980)^+$, $\Xi_c(3077)^+$, and $\Xi_c(3077)^0$, with an improvement over existing measurements of the $\Xi_c(2980)^+$ mass and width.
For the three-body final states, we also search for decays through intermediate resonant $\Sigma_c(2455)^{++}K^-$, $\Sigma_c(2455)^{0}\KS$, $\Sigma_c(2520)^{++}K^-$, and $\Sigma_c(2520)^{0}\KS$ channels.
The $\Sigma_c(2455)$ and $\Sigma_c(2520)$ baryons decay exclusively to $\Lambda_c^+\pi$.
We find evidence for two additional new states $\Xi_c(3055)^+$ and $\Xi_c(3123)^+$ decaying through the intermediate-resonant channels $\Sigma_c(2455)^{++}K^-$ and $\Sigma_c(2520)^{++}K^-$, respectively.
We measure the mass, natural width, yield, and the product of production cross section and decay branching fractions when there is evidence for an excited charm-strange baryon.
Also, where applicable, the intermediate resonant decay fractions are measured.

\section{Analysis Overview}

This analysis relies primarily on the charged-particle tracking and 
particle-identification capabilities of the \babar\ detector.
A detailed description of the \babar\ detector is presented in Ref.~\cite{babar}.
The charged-particle tracking system consists of a five-layer double-sided silicon vertex tracker (SVT) and a 40-layer drift chamber (DCH).
Discrimination between charged pions, kaons, and protons relies on ionization energy loss (\dedx) in the DCH and SVT, and on Cherenkov photons detected in a ring-imaging detector (DIRC).
A CsI(Tl) crystal calorimeter is used to identify electrons and photons.
These four detector subsystems are mounted inside a 1.5-T solenoidal superconducting magnet.
The instrumented flux return for the solenoidal magnet provides muon identification.

We produce samples of simulated events using the Monte Carlo (MC) generators JETSET74 \cite{jetset} and EVTGEN~\cite{EVTGEN} with a full detector simulation based on GEANT4~\cite{MC}.
We produce about four million simulated $\epem\rightarrow c\bar{c}$ events in which at least one of the primary charm quarks hadronizes into an excited charm-strange baryon that decays according to one of the studied decay channels.
All particle decays are generated according to phase space.
Reconstruction efficiencies are estimated based on the excited charm-strange baryon kinematic distributions from JETSET74.
The simulated samples are also used to estimate measured invariant-mass resolutions.

In data, we study the invariant-mass region between 2.91\gevcc and 3.15\gevcc for potential $\Xi_c$ states.  
We do not examine candidates in this region while optimizing the selection criteria in order to minimize potential experimenters' bias. 
Selection criteria are chosen to maximize the expected significance of signal events averaged over five reconstructed $\Lambda_c^+$ decay modes:
$pK^-\pi^+$, $p\KS$, $p\KS\pi^-\pi^+$, $\Lambda\pi^+$, and $\Lambda\pi^+\pi^-\pi^+$.
The expected significance of each decay mode is estimated based on the efficiency measured with samples of simulated signal events, and the number of candidates in upper (above 3.15\gevcc) and lower (below 2.91\gevcc) $\Xi_c$ invariant-mass side-bands in data.

The selection criteria are based on proton, kaon, and pion particle identification, among several other reconstruction parameters.
Reconstructed tracks in the entire candidate decay chain are simultaneously fit with vertex constraints and with the $\Lambda_c^+$, $\Lambda$, and $\KS$ candidate masses constrained to their world-average values~\cite{PDG}.
The total $\chi^2$ probability of the fitted track vertices in the decay chain is required to be greater than 1\%.
The vertex-constrained $\Lambda_c^+$ mass (with no mass constraint) is required to be within 1.75 times the mass resolution ($\sim 5\mevcc$) of the world-average $\Lambda_c^+$ mass~\cite{PDG}.
The vertex-constrained $\Lambda$ and $\KS$ masses are required to be within 6.5\mevcc and 10.0\mevcc, respectively, of their world-average values~\cite{PDG}.
The measured $\Lambda$ and $\KS$ flight lengths divided by their errors are required to be greater than $2.5$ and $3.0$, respectively.
The momentum of the $\Xi_c$ candidate in the \epem center-of-mass frame ($p^*$) is required to be above 2.9\gevc.
This $p^*$ requirement significantly reduces background from random combinations of tracks.

For each $\Xi_c$ decay mode, the selected candidates in data are divided into five samples based on the reconstructed $\Lambda_c^+$ decay mode.
The signal yields in each sample are extracted with an extended, unbinned, maximum-likelihood fit.
All five samples are fit simultaneously.
The signal probability density functions (PDF) share the values of parameters that describe the signal shape and, for three-body final states, the intermediate-resonant decay fractions.
For $\Lambda_c^+$ decay mode $m$, the corresponding portion of the likelihood function $L_m$ has the form
\begin{equation}
\label{eq:like1}
L_m= \prod_i^{N_m}\left[\sum_{\eta=1}^{N_s}S_m^{\eta}P_s^{\eta}(\vec{a};x_m^i) + \sum_{\kappa=1}^{N_b}B_m^{\kappa}P_b^{\kappa}(\vec{b};x_m^i)\right]\,,
\end{equation}
where $S_m^{\eta}$ is the number of candidates in signal-component $\eta$,
$B_m^{\kappa}$ is the number of candidates in background-component $\kappa$,
$\vec{a}$ and $\vec{b}$ are the shape parameters for the $N_s$ signal-PDF components ($P_s^{\eta}$) and the $N_b$ background-PDF components ($P_b^{\kappa}$),
and $x_m^i$ represents the measured masses for the $N_m$ candidates in decay mode $m$.
The mass, natural width, and resonant fraction results listed in Table~\ref{tb:results3body} are the measured shape parameters $\vec{a}$ for the respective signal PDFs. 
Combining the five $L_m$, the full extended likelihood function ${\cal L}$ has the form
\begin{equation}
\label{eq:like2}
{\cal L}=\exp\left[-\sum_{m=1}^{5}(N_m-\sum_{\eta}^{N_s} S_m^{\eta}-\sum_{\kappa}^{N_b} B_m^{\kappa})\right]\times\prod_{m=1}^{5}L_m\,.
\end{equation}
Each parameter is allowed to vary over a range that is large enough so that it does not constrain any of the final results;  
in particular, the number of signal candidates is allowed to be negative.
For each signal-PDF component ($P^{\eta}_s$), the numbers of measured candidates with $\Lambda_c^+$ decay mode $m$ ($S_m^{\eta}$) are summed together to determine the total yields listed in Table~\ref{tb:results3body} or to determine the products of cross-sections and branching fractions listed in Tables~\ref{tb:3bodCross} and \ref{tb:24bodCross}.

Statistically significant signals for excited charm-strange baryons are observed only in the analysis of three-body final states.
For the two-body and four-body final states, we only search for the baryons observed in the three-body final states.
The $\Xi_c$ masses and widths measured with the three-body candidates are used as Gaussian constraints in the fits to samples of two-body and four-body candidates by multiplying the likelihood function in Eq.~(\ref{eq:like2}) by
\begin{equation}
\label{eq:Gauss}
\exp\left[-\frac{({\cal M}-\mu)^2}{2\sigma_{\cal M}^2}-\frac{(\Gamma-\gamma)^2}{2\sigma_{\Gamma}^2}\right]\,,
\end{equation}
where ${\cal M}$ is the mass and $\Gamma$ is the natural width of the $\Xi_c$ state measured from the fit to three-body candidates, $\sigma_{\cal M}$ and $\sigma_{\Gamma}$ are the corresponding uncertainties, and $\mu$ and $\gamma$ are the constrained signal mean and width parameters.

\section{\boldmath Decays to $\Lambda_c^+K^-\pi^+$ and $\Lambda_c^+\KS\pi^-$}
The studies of the three-body final states $\Lambda_c^+K^-\pi^+$ and $\Lambda_c^+\KS\pi^-$ are based on fits to the two-dimensional invariant-mass distributions
$M(\Lambda_c^+K^-\pi^+)$ versus $M(\Lambda_c^+\pi^+)$, and $M(\Lambda_c^+\KS\pi^-)$ versus $M(\Lambda_c^+\pi^-)$.
The experimental resolution for these two- and three-body invariant masses
varies from about 1.0\mevcc to 2.5\mevcc. 
With these two-dimensional mass distributions, we can incorporate the intermediate resonances $\Sigma_c(2455)^{++,0}$ and $\Sigma_c(2520)^{++,0}$ in the fits.
No other known intermediate resonances are kinematically allowed.
We fit a region defined by the kinematically allowed mass thresholds, and  $\Lambda_c^+\KS\pi^-$ and $\Lambda_c^+K^-\pi^+$ invariant masses up to $3150\mevcc$.
Two-dimensional mass plots for candidates in the fit region are shown in Fig.~\ref{fg:3K}(a) for $\Lambda_c^+K^-\pi^+$ candidates and in Fig.~\ref{fg:3Ks}(a) for $\Lambda_c^+\KS\pi^-$ candidates. 

For the two-dimensional maximum-likelihood fit we use a PDF with background components ($P_b$) and signal components ($P_s$) used to describe four candidate categories:
resonant combinatoric background, non-resonant combinatoric background, resonant signal, and non-resonant signal.
We use a double-Voigtian resonance shape (a non-relativistic Breit-Wigner distribution convolved with a resolution function that is the sum of two Gaussian functions) to describe each peak in invariant mass.
These peaks include the $\Xi_c$ states as well as the $\Sigma_c$ states associated with both $\Xi_c$ decays and backgrounds.
The double-Voigtian resonance shape does not account for possible interference effects due to overlapping resonances. 
All resolution parameters are fixed to values measured in the MC samples.
The masses and widths of the $\Sigma_c(2455)^{++}$ and $\Sigma_c(2520)^{++}$ are free parameters in the fits, while
the masses and widths of the $\Sigma_c(2455)^{0}$ and $\Sigma_c(2520)^{0}$ are fixed to the world average values~\cite{PDG} because of the limited statistical sensitivity of the data to these states.
The double-Voigtian resonance shape that accounts for excited charm-strange baryon decays through an intermediate $\Sigma_c$ resonance is multiplied by a two-body phase-space factor,
and the double-Voigtian resonance shape that accounts for direct three-body decays is multiplied by a three-body phase-space factor.
The non-$\Sigma_c$ backgrounds are modeled with a threshold function proportional to
\begin{equation}
\label{eq:thresh}
M\left[\left( \frac{M}{t} \right)^{2}-1 \right]^{\alpha}\exp\left[\beta(\left( \frac{M}{t} \right)^{2}-1) \right]\,,
\end{equation}
where $M$ is the mass variable in the distribution of which there is a minimum kinematic threshold, $t$ is the value of the threshold, and $\alpha$ and $\beta$ are shape parameters.
The free parameters ($\vec{a}$) of the signal-PDF components include the masses, natural widths, and resonant fractions of the excited charm-strange baryon signals.

\subsection{\boldmath $\Lambda_c^+K^-\pi^+$ Results}
\begin{figure}
\begin{minipage}[t]{0.5\textwidth}
\includegraphics[width=.95\textwidth]{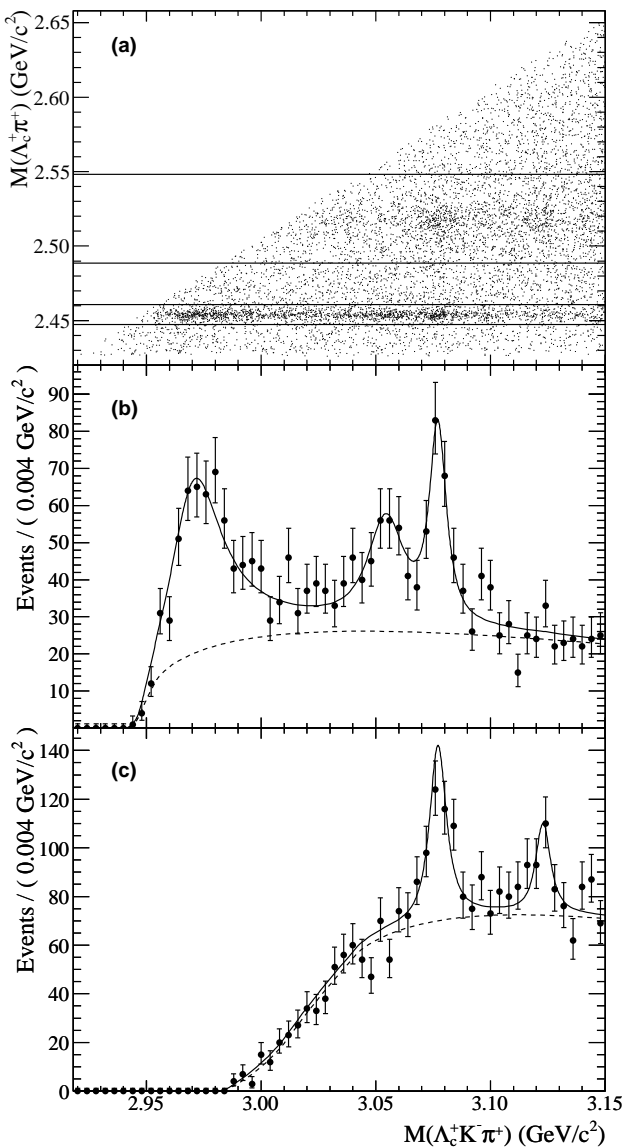}
\caption{(a) Scatter plot of $M(\Lambda_c^+\pi^+)$ versus $M(\Lambda_c^+K^-\pi^+)$, and (b,c) projections of the $M(\Lambda_c^+K^-\pi^+)$ distributions, in two ranges of $M(\Lambda_c^+\pi^+)$, for data (points with error bars) summed over all five reconstructed $\Lambda_c^+$ decay modes combined.  
The curves correspond to the two-dimensional fit results.
The $M(\Lambda_c^+\pi^+)$ ranges are (b) within 3.0 natural widths of the $\Sigma_c(2455)^{++}$ mass, and (c) within 2.0 natural widths of the $\Sigma_c(2520)^{++}$ mass.
These $M(\Lambda_c^+\pi^+)$ ranges are delineated by the horizontal lines on the scatter plot in (a).
The projections of the combined background PDF components are illustrated by the dashed curves.}
\label{fg:3K}
\end{minipage}
\end{figure}

Projections of the $\Lambda_c^+K^-\pi^+$ invariant-mass distribution, for all five reconstructed $\Lambda_c^+$ decay modes combined, are shown in Figs.~\ref{fg:3K}(b) and (c) with the combined components of the fit shown.
The fit allows for four excited charm-strange baryon states; the masses and widths are free parameters in the fit. 
The data and fit projections correspond to $M(\Lambda_c^+\pi^+)$ ranges within 3.0 natural widths of the $\Sigma_c(2455)^{++}$ mass (Fig.~\ref{fg:3K}(b)) and 2.0 natural widths of the $\Sigma_c(2520)^{++}$ mass (Fig.~\ref{fg:3K}(c))~\cite{PDG}.
These $M(\Lambda_c^+\pi^+)$ ranges are delineated by the horizontal lines in Fig.~\ref{fg:3K}(a).
A two-dimensional ($M(\Lambda_c^+\pi^+)$ and $M(\Lambda_c^+K^-\pi^+)$) binned $\chi^2$ probability of 88\% is calculated for the fit using roughly 300 equally sized bins each containing over 10 candidates.
Four excited charm-strange states are observed at the approximate $M(\Lambda_c^+K^-\pi^+)$ invariant-masses of 2970\mevcc, 3055\mevcc, 3077\mevcc, and 3123\mevcc.
We use the change in maximum ln-likelihood value when a signal is removed from the fit, and the number of free parameters in each signal PDF component, to estimate statistical significances of \mbox{$>9.0\,\sigma$} ($\Delta\ln {\cal L}=81$), $6.4\,\sigma$ ($\Delta\ln {\cal L}=31$), \mbox{$>9.0\,\sigma$} ($\Delta\ln {\cal L}=91$), and $3.6\,\sigma$ ($\Delta\ln {\cal L}=15$) for each state, in order of increasing mass.
These estimated significances do not statistically account for fluctuations in background level across the entire range of invariant masses over which the study was conducted. 
This means that these significances do not correspond to those of a search.
Rather, they correspond to significances of signals incorporated into the PDF after evidence for the corresponding states had been observed.

The $\Xi_c(2980)^+$ PDF includes a nonresonant $\Lambda_c^+K^-\pi^+$ component and a $\Sigma_c(2455)^{++}K^-$ component. 
(The $\Sigma_c(2520)^{++}K^-$ final state is not kinematically allowed for the $\Xi_c(2980)^+$.)
For the $\Xi_c(2980)^+$, $(55\pm7\pm13)$\% of the signal is found to be due to decays through  $\Sigma_c(2455)^{++}K^-$.
(The uncertainties are statistical and systematic. The estimation of systematic errors is discussed below.)
The $\Xi_c(3077)^+$ PDF includes a nonresonant $\Lambda_c^+K^-\pi^+$ component, a $\Sigma_c(2455)^{++}K^-$ component, and a $\Sigma_c(2520)^{++}K^-$ component.
For the $\Xi_c(3077)^+$, $(95\pm14\pm6)$\% of the signal is found to be due to decays through intermediate resonances.
A lower limit of 80\% for this fraction is determined, at 90\% confidence level, from a numerical integration of the posterior probability density calculated from the likelihood distribution and a zero prior for fractions above 100\%.
Of these intermediate-resonant decays, $(45\pm5\pm5)$\% are through $\Sigma_c(2455)^{++}K^-$.
Separate fits to the data that include $\Lambda_c^+K^-\pi^+$, $\Sigma_c(2455)^{++}K^-$, and $\Sigma_c(2520)^{++}K^-$ PDF components for $\Xi_c(3055)^+$ and $\Xi_c(3123)^+$ signal indicate that these states only have signals for decays through the intermediate-resonant states $\Sigma_c(2455)^{++}K^-$ and $\Sigma_c(2520)^{++}K^-$, respectively.
For all measurements, the $\Xi_c(3055)^+$ and $\Xi_c(3123)^+$ PDFs include only these intermediate-resonant components.

The newly identified $\Xi_c(3055)^+$ and $\Xi_c(3123)^+$ baryons are found to have statistically significant signals with
widths that are larger in value than our mass resolution,
$p^*$ values distributed toward higher momenta than the combinatoric background,
and consistent signal decay rates between the five $\Lambda_c^+$ final states. 
Four additional samples are studied for possible sources of peaking background that could be mistaken for any of the excited charm-strange baryon signals.
These four samples are a ``wrong-sign'' $\Lambda_c^+K^+\pi^-$ data sample, a $\Lambda_c^+$ mass-sideband data sample, a MC sample of $\epem\rightarrow c\overline{c}$ events, and a data sample in which the $\Lambda_c^+K^-\pi^+$ invariant mass is recalculated substituting a pion mass for the kaon mass ($\Lambda_c^+\pi^-\pi^+$).
The last data sample would reveal excited $\Lambda_c^+$ decays in which the $\pi^-$ has been misidentified as a $K^-$; no evidence for such decays is found.
We also determined that misidentifed excited $\Lambda_c^+$ decays result in peaks that are asymmetric in mass and are much broader than the signals observed in data.
The wrong-sign data sample could show evidence of peaking backgrounds due to $\Sigma_c^0$ decays that are reconstructed with an additional kaon, but no such evidence is found. 
The $\Lambda_c^+$ mass-sideband data sample and the generic-$c\overline{c}$ MC sample are searched for unexpected sources of peaking background.
No sources of peaking background are found. 
We conclude from these studies that the signals for $\Xi_c(2980)^+$, $\Xi_c(3055)^+$, $\Xi_c(3077)^+$, and $\Xi_c(3123)^+$ are not peaking backgrounds.
These studies also confirm that the threshold function in Eq.~(\ref{eq:thresh}) accurately models non-peaking background shapes.

\subsection{\boldmath $\Lambda_c^+\KS\pi^-$ Results} 
\begin{figure}
\begin{minipage}[t]{0.5\textwidth}
\includegraphics[width=.95\textwidth]{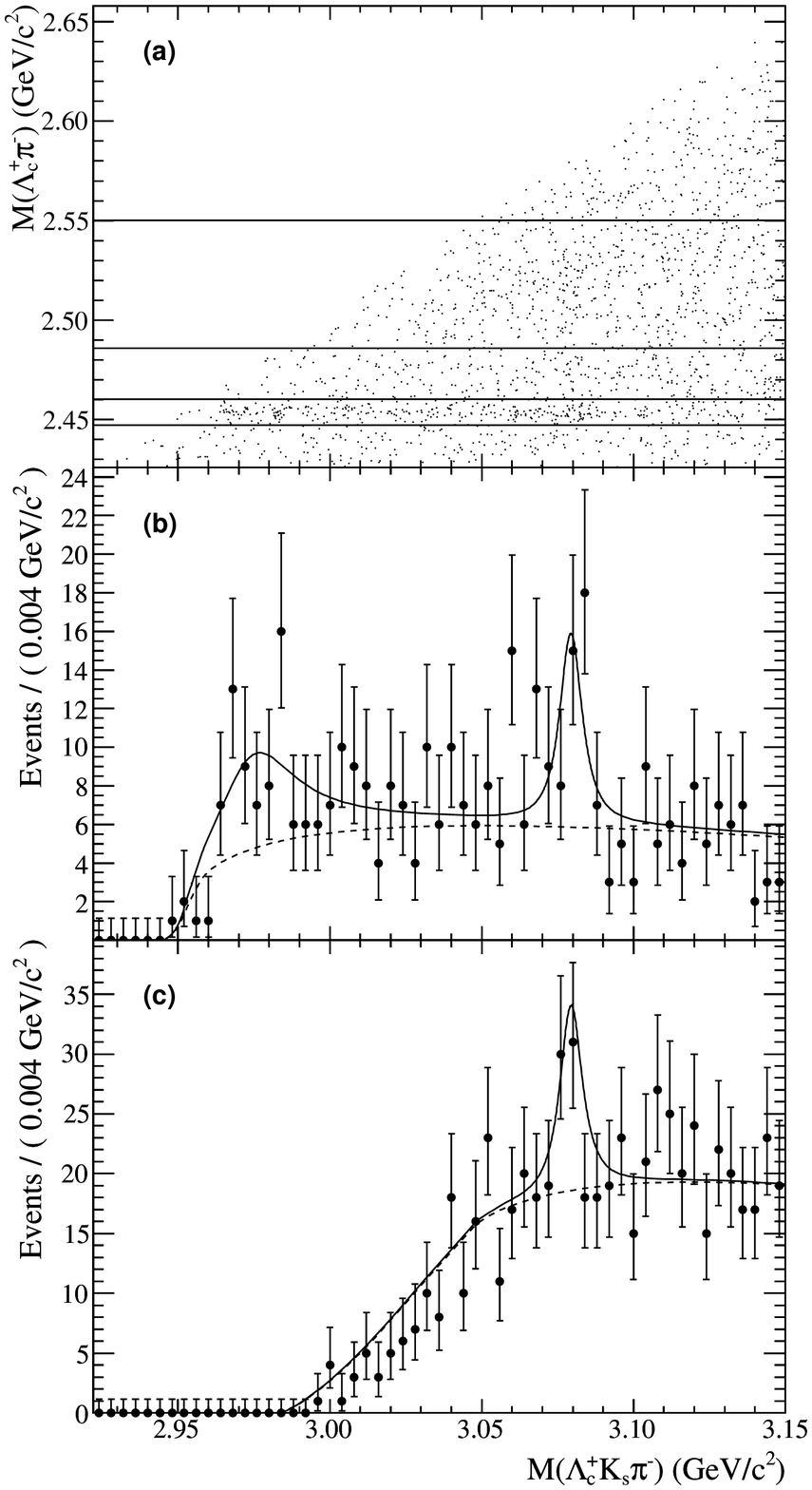}
\caption{(a) Scatter plot of $M(\Lambda_c^+\pi^-)$ versus $M(\Lambda_c^+\KS\pi^-)$, and (b,c) projections of the $M(\Lambda_c^+\KS\pi^-)$ distribution, in two ranges of $M(\Lambda_c^+\pi^-)$, for data (points with error bars) summed over all five reconstructed $\Lambda_c^+$ decay modes combined. The curves correspond to the two-dimensional fit results.
The $M(\Lambda_c^+\pi^-)$ ranges are (b) within 3.0 natural widths of the $\Sigma_c(2455)^0$ mass, and (c) within 2.0 natural widths of the $\Sigma_c(2520)^0$ mass.
These $M(\Lambda_c^+\pi^-)$ ranges are delineated by the horizontal lines on the scatter plot in (a). 
The projections of the combined background PDF components are illustrated by the dashed curves.}
\label{fg:3Ks}
\end{minipage}
\end{figure}

Projections of the $\Lambda_c^+\KS\pi^-$ invariant-mass distribution, for all five reconstructed $\Lambda_c^+$ decay modes combined, are shown in Figs.~\ref{fg:3Ks}(b) and (c).
Projections of the combined compenents of the fit are also shown.
The data and fit projections correspond to $M(\Lambda_c^+\pi^-)$ ranges within 3.0 natural widths of the $\Sigma_c(2455)^0$ mass (Fig.~\ref{fg:3Ks}(b))and 2.0 natural widths of the $\Sigma_c(2520)^0$ mass (Fig.~\ref{fg:3Ks}(c))~\cite{PDG}.
These $M(\Lambda_c^+\pi^-)$ ranges are delineated by the horizontal lines in Fig.~\ref{fg:3Ks}(a).
The fit shown in Fig.~\ref{fg:3Ks} accounts for two excited charm-strange baryons: $\Xi_c(2980)^0$ and $\Xi_c(3077)^0$.
A two-dimensional binned $\chi^2$ probability of 62\% is calculated (in the same way as described for the fit to $\Lambda_c^+K^-\pi^+$ data) for the fit when both the $\Xi_c(2980)^0$  and $\Xi_c(3077)^0$ PDF components are included,
while a 54\% $\chi^2$ probability is calculated when only the $\Xi_c(3077)^0$ PDF component is included.
Estimated statistical significances of $1.7\,\sigma$ ($\Delta\ln {\cal L}=8$) and $4.5\,\sigma$ ($\Delta\ln {\cal L}=21$) are found for the $\Xi_c(2980)^0$ and $\Xi_c(3077)^0$ states, respectively, based on changes in ln-likelihood values and the number of free parameters in each signal PDF component.
Despite the low statistical significance of the $\Xi_c(2980)^0$ signal, we use the prior assumption that this state exists as the isospin partner of the $\Xi_c(2980)^+$ and include the $\Xi_c(2980)^0$ PDF components for all measurements based on the sample of $\Lambda_c^+\KS\pi^-$ candidates.

The $\Xi_c(2980)^0$ PDF includes both a nonresonant $\Lambda_c^+\KS\pi^-$ component and an intermediate-resonant $\Sigma_c(2455)^{0}\KS$ component.
The $\Xi_c(3077)^0$ PDF includes a nonresonant $\Lambda_c^+\KS\pi^-$ component, and intermediate-resonant $\Sigma_c(2455)^{0}\KS$ and $\Sigma_c(2520)^{0}\KS$ components;
$(78\pm21\pm5)$\% of the $\Xi_c(3077)^0$ signal is found to decay through the intermediate resonances.
Of this intermediate-resonant signal, $(44\pm12\pm7)$\% is $\Sigma_c(2455)^{++}K^-$.
No statistically significant signals are found for $\Xi_c(3055)^0$ or $\Xi_c(3123)^0$ states (the neutral isospin partners of $\Xi_c(3055)^+$ and $\Xi_c(3123)^+$).

\subsection{Systematic Uncertainties}
Several sources of systematic uncertainty are investigated and quantified for the measurements of charm-strange baryons decaying to $\Lambda_c^+K^-\pi^+$ and $\Lambda_c^+\KS\pi^-$ final states;
they are listed in Table~\ref{tb:sys3bod}.
Uncertainties in PDF modeling are estimated from additional fits to the data with different mass resolution parameters, background shapes, and phase-space threshold masses.
Because of their relatively low statistical significances, we include the effects of excluding a $\Xi_c(2980)^0$ or $\Xi_c(3123)^+$ PDF component in the relevant fits.
Systematic uncertainties on the measured mass, associated with SVT misalignments, angular dependence of tracking performance, energy loss in detector material, the magnetic field, and material magnetization, were extensively studied in \babar\ for a precision measurement of the $\Lambda_c^+$ mass~\cite{lambdac}.
Since the $Q$-values of the excited charm-strange baryon decays in this analysis are similar to those of the $\Lambda_c^+$ decays used in the mass measurement, we assign the same systematic error of 0.14\mevcc.
This uncertainty is much smaller than those from other sources.

The systematic uncertainties are added in quadrature and are listed along with the measurements of masses, widths, yields, and intermediate-resonant decay fractions in Table~\ref{tb:results3body}.
To determine the effect of systematic uncertainties on the estimated significance of the $\Xi_c(3055)^+$ and $\Xi_c(3123)^+$ states, the change in ln-likelihood is recalculated with the same PDF modeling changes described in the previous paragraph.
Only for the $\Xi_c(3123)^+$ do any of these changes reduce the significance of the signal; 
the modified background shapes reduce the significance from 3.6 to $3.0\,\sigma$ ($\Delta\ln {\cal L}=12$).
This lower significance is listed in parentheses in Table~\ref{tb:results3body}.

\begin{table}
\caption{\label{tb:results3body} Measured masses, widths, yields, resonant decay fractions, and significances for $\Xi_c(2980)^+$, $\Xi_c(3055)^+$, $\Xi_c(3077)^+$, and $\Xi_c(3123)^+$ baryons decaying to $\Lambda_c^+K^-\pi^+$, and $\Xi_c(2980)^0$ and $\Xi_c(3077)^0$ baryons decaying to $\Lambda_c^+\KS\pi^-$.
The first errors are statistical and the second errors are systematic.
All signal candidates are required to have $p^*$ greater than 2.9\gevc.
The rows labeled ``Resonant'' give the fraction of each signal that decays through intermediate resonances.
The $\Xi_c(3077)^+$ resonant fraction is a 90\% confidence-level lower limit.
The row labeled ``$\Sigma_c(2455)$'' gives the fraction of resonant decays that proceed through $\Sigma_c(2455)K$.
The only state for which systematic uncertainties reduce the calculated significance at the stated precision is the $\Xi_c(3123)^+$; the lower significance is given in parentheses.
}
\begin{center}
\begin{tabular}{lcccc} \hline
               &\,& $\Xi_c(2980)^+$     &\,& $\Xi_c(2980)^0$        \\ \hline\hline
Mass (\mevcc)  &\,& $2969.3\pm2.2\pm1.7$&\,& $2972.9\pm4.4\pm1.6$   \\
Width (\mev)   &\,& $27\pm8\pm2$  &\,& $31\pm7\pm8$           \\
Yield          &\,& $756\pm178\pm104$   &\,& $67\pm33\pm29$         \\
Resonant (\%)  &\,& $55\pm7\pm13$       &\,& |         \\
Significance   &\,& $>9.0\,\sigma$      &\,& $1.7\,\sigma$          \\ \hline
               &\,& $\Xi_c(3077)^+$     &\,& $\Xi_c(3077)^0$        \\ \hline\hline
Mass (\mevcc)  &\,& $3077.0\pm0.4\pm0.2$&\,& $3079.3\pm1.1\pm0.2$   \\
Width (\mev)   &\,& $5.5\pm1.3\pm0.6$   &\,& $5.9\pm2.3\pm1.5$      \\
Yield          &\,& $403\pm54\pm27$     &\,& $90\pm22\pm15$         \\
Resonant (\%)  &\,& $>80$       &\,& $78\pm21\pm5$          \\
$\Sigma_c(2455)$ (\%)&\,&$45\pm5\pm5$   &\,& $44\pm12\pm7$          \\
Significance   &\,& $>9.0\,\sigma$      &\,& $4.5\,\sigma$          \\ \hline
               &\,& $\Xi_c(3055)^+$     &\,& $\Xi_c(3123)^+$        \\ \hline\hline
Mass (\mevcc)  &\,& $3054.2\pm1.2\pm0.5$&\,& $3122.9\pm1.3\pm0.3$   \\
Width (\mev)   &\,& $17\pm6\pm11$       &\,& $4.4\pm3.4\pm1.7$      \\
Yield          &\,& $218\pm53\pm79$     &\,& $101\pm34\pm9$         \\
Significance   &\,& $6.4\,\sigma$       &\,& $3.6\,\sigma$~$(3.0\,\sigma)$  \\ \hline\hline
\end{tabular}
\end{center}
\end{table}

\begin{table}
\caption{\label{tb:sys3bod}
Systematic uncertainties on the masses, widths, yields, resonant fractions ($R_1$), and $\Sigma_c(2455)$ resonant fractions ($R_2$) determined from the samples of $\Lambda_c^+ K^-\pi^+$ and $\Lambda_c^+ \KS \pi^-$ candidates.
Uncertainties associated with mass resolution, background-PDF shapes, phase-space thresholds, the inclusion of the $\Xi_c(3123)^+$ or $\Xi_c(2980)^0$ signal shapes, and the mass scale are listed.
The systematic errors from each source are added in quadrature.
NA indicates that a source of systematic uncertainty is not applicable.}
\begin{center}
\begin{tabular}{lccccc} \hline
                    & Mass      & Width    & Yield    & $R_1$     & $R_2$  \\ 
                    & (\mevcc)  & (\mev)   & (\%)     & (\%)      & (\%)   \\ \hline\hline
$\Xi_c(2980)^+$ \\
Mass Resolution     & $\pm0.6$  & $\pm1.3$ & $\pm\p4$ & $\pm\p4$  & NA \\
Background Shape    & $\pm1.0$  & $\pm0.3$ & $\pm\p8$ & $\pm\p8$  & NA \\
Phase-Space Thresh. & $\pm1.2$  & $\pm0.3$ & $\pm10$  & $\pm\p9$  & NA \\
Signal Inclusion    & $\pm0.4$  & $\pm0.4$ & $\pm\p3$ & $\pm\p3$  & NA \\
Mass Scale          & $\pm0.1$  & NA       & NA       & NA        & NA \\ \hline
Total               & $\pm1.7$  & $\pm1.5$ & $\pm14$  & $\pm13$ & NA \\ \hline
$\Xi_c(3055)^+$ \\
Mass Resolution     & $\pm0.4$  & $\pm7.3$ & $\pm26$  & NA     & NA \\
Background Shape    & $\pm0.3$  & $\pm6.8$ & $\pm22$  & NA     & NA \\
Phase-Space Thresh. & $\pm0.1$  & $\pm4.2$ & $\pm11$  & NA     & NA \\
Signal Inclusion    & $\pm0.0$  & $\pm1.4$ & $\pm\p5$ & NA     & NA \\
Mass Scale          & $\pm0.1$  & NA       & NA       & NA     & NA \\ \hline
Total               & $\pm0.5$  & $\pm11$  & $\pm36$  & NA     & NA \\ \hline
$\Xi_c(3077)^+$ \\
Mass Resolution     & $\pm0.11$ & $\pm0.4$ & $\pm3.2$ & $\pm3.6$  & $\pm2.5$ \\
Background Shape    & $\pm0.10$ & $\pm0.2$ & $\pm0.1$ & $\pm1.2$  & $\pm2.7$ \\
Phase-Space Approx. & $\pm0.09$ & $\pm0.3$ & $\pm3.0$ & $\pm1.4$  & $\pm1.9$ \\
Signal Inclusion    & $\pm0.05$ & $\pm0.2$ & $\pm5.0$ & $\pm4.6$  & $\pm2.4$ \\
Mass Scale          & $\pm0.14$ & NA       & NA       & NA        & NA       \\ \hline
Total               & $\pm0.18$ & $\pm0.6$ & $\pm6.7$ & $\pm6.1$  & $\pm4.8$ \\ \hline
$\Xi_c(3123)^+$ \\
Mass Resolution     & $\pm0.3$  & $\pm1.5$ & $\pm5.0$   & NA     & NA \\
Background Shape    & $\pm0.2$  & $\pm0.6$ & $\pm6.9$   & NA     & NA \\
Phase-Space Thresh. & $\pm0.1$  & $\pm0.5$ & $\pm3.0$   & NA     & NA \\
Mass Scale          & $\pm0.1$  & NA       & NA         & NA     & NA \\ \hline
Total               & $\pm0.3$  & $\pm1.7$ & $\pm8.9$   & NA     & NA \\ \hline
$\Xi_c(2980)^0$ \\
Mass Resolution     & $\pm0.6$  & $\pm7.1$ & $\pm16$   & $\pm\p1$ & NA \\
Background Shape    & $\pm1.3$  & $\pm3.7$ & $\pm37$   & $\pm14$  & NA \\
Phase-Space Thresh. & $\pm0.7$  & $\pm1.7$ & $\pm12$   & $\pm17$  & NA \\
Mass Scale          & $\pm0.1$  & NA       & NA        & NA       & NA \\ \hline
Total               & $\pm1.6$  & $\pm8.2$ & $\pm42$   & $\pm22$  & NA \\ \hline
$\Xi_c(3077)^0$ \\
Mass Resolution     & $\pm0.01$ & $\pm0.3$ & $\pm\p1$  & $\pm0.4$ & $\pm0.3$ \\
Background Shape    & $\pm0.12$ & $\pm0.2$ & $\pm\p1$  & $\pm4.1$ & $\pm3.1$ \\
Phase-Space Thresh. & $\pm0.03$ & $\pm0.1$ & $\pm\p1$  & $\pm2.4$ & $\pm0.2$ \\
Signal Inclusion    & $\pm0.02$ & $\pm1.5$ & $\pm17$   & $\pm0.9$ & $\pm6.3$ \\
Mass Scale          & $\pm0.14$ & NA       & NA        & NA       & NA       \\ \hline
Total               & $\pm0.19$ & $\pm1.5$ & $\pm17$   & $\pm4.9$ & $\pm7.0$ \\ \hline
\end{tabular}
\end{center}
\end{table}

\subsection{Products of Cross-Section and Branching Fractions} 
Reconstruction efficiencies are estimated from JETSET74 simulations of the \epem production of excited charm-strange baryons.
The efficiencies are studied as a function of the kinematic variables 
$M(\Lambda_c^+ K)^2$, $M(K\pi^\pm)^2$, and $M(\Lambda_c^+\pi^\pm)^2$, where $K$ denotes $K^-$ or $\KS$.
A slight 6\% relative variation in efficiency is found only at large values of $M(\Lambda_c^+\pi^+)^2$.
Therefore, we use an efficiency for each $\Lambda_c^+ K^- \pi^+$ candidate that depends on the measured value of $M(\Lambda_c^+\pi^+)^2$.
Reconstruction efficiencies for states that decay through $\Lambda_c^+K^-\pi^+$ range from about 1\% to 9\%, depending on the reconstructed $\Lambda_c^+$ decay mode.
Reconstruction efficiencies for states that decay through $\Lambda_c^+\KS\pi^-$ range from about 0.4\% to 3.2\%.

For each excited charm-strange baryon state, the number of baryons produced in the \babar detector that decay to $\Lambda_c^+K^-\pi^+$ or $\Lambda_c^+\KS\pi^-$, where $\Lambda_c^+\rightarrow pK^-\pi^+$, is estimated using the signal yields for all five $\Lambda_c$ decay modes, the reconstruction efficiencies, the $\Lambda_c^+$ branching-fraction ratios~\cite{PDG}, and a ``Best Linear Unbiased Estimate'' (BLUE) method~\cite{BLUE}.
The BLUE method accounts for correlated errors between the five signal yields from the five reconstructed $\Lambda_c^+$ decay modes, as well as correlated errors between the five estimated efficiencies. 
The estimated number of excited charm-strange baryons produced with $p^*>2.9\gevc$ and the integrated luminosity are used to calculate the product of cross-section and branching fractions for each excited charm-strange baryon:
\begin{equation}
\label{eq:cross}
\sigma(\epem\rightarrow\Xi_c^*X){\cal B}(\Xi_c^*\rightarrow Y){\cal B}(\Lambda_c^+\rightarrow pK^-\pi^+)\,,
\end{equation}
where $\Xi_c^*$ is an excited charm-strange baryon, $X$ is the unreconstructed portion of the $\epem$ event, and $Y$ is $\Lambda_c^+K^-\pi^+$ or $\Lambda_c^+\Kzb\pi^-$.
With these calculations, it is assumed that 34.6\% of the $\Kzb$ mesons decay as $\KS\rightarrow\pi^+\pi^-$~\cite{PDG}.
The measured products of cross-section and branching fractions are listed in Table~\ref{tb:3bodCross}.

Upper limits at 90\% confidence level are calculated on the product of cross-section and branching fractions for the neutral states $\Xi_c(2980)^0$, $\Xi_c(3055)^0$, and $\Xi_c(3123)^0$.
The upper limits are determined from numerical integration of posterior probability densities calculated from the likelihood distributions, with a uniform positive prior for the product of cross-section and branching fractions greater than zero and a zero prior below.
Gaussian constraints on efficiencies and $\Lambda_c^+$ branching-fraction ratios are included in the likelihood functions;
these modified likelihood functions constrain the ratios of fitted yields from each $\Lambda_c^+$ decay mode and incorporate systematic uncertainties from these ratios of yields.
Also, in calculating each upper limit, the signal shape parameters of the baryon are given Gaussian constraints based on the measured values in Table~\ref{tb:results3body}.
For $\Xi_c(3055)^0$ and $\Xi_c(3123)^0$ parameters, we use the values of the parameters measured for their isospin partners. 
All Gaussian constraints involve a factor analogous to that in Eq.~(\ref{eq:Gauss}).
The upper limits on the products of cross-section and branching fractions are listed in Table~\ref{tb:3bodCross}.

\begin{table}
\caption{\label{tb:3bodCross} The products of cross-section and branching fractions (Eq.~(\ref{eq:cross})) for excited charm-strange baryons produced with $p^*>2.9\gevc$ and decaying to three-body final states.
The first errors are statistical and the second errors are systematic.
Upper limits are at 90\% confidence level.
}
\begin{center}
\begin{tabular}{ccccc} \hline
$\Xi_c^*\rightarrow$ &\,& $\Lambda_c^+K^-\pi^+$     &\,& $\Lambda_c^+\Kzb\pi^-$ \\ \hline\
$\Xi_c(2980)$        &\,& $(11.8\pm3.4\pm2.2)$\,fb  &\,& $<15$\,fb                        \\
$\Xi_c(3055)$        &\,& $\p(2.2\pm1.2\pm0.7)$\,fb &\,& $<\p7$\,fb                       \\
$\Xi_c(3077)$        &\,& $\p(8.1\pm1.2\pm0.8)$\,fb &\,& $(6.2\pm2.1\pm1.5)$\,fb          \\
$\Xi_c(3123)$        &\,& $\p(1.6\pm0.6\pm0.2)$\,fb &\,& $<1.4$\,fb                       \\ \hline\hline
\end{tabular}
\end{center}
\end{table}

\section{\boldmath Decays to $\Lambda_c^+\KS$ and $\Lambda_c^+K^-$}
\label{sec:2bod}

The studies of decays to the two-body final states $\Lambda_c^+\KS$ and $\Lambda_c^+K^-$ are based on one-dimensional fits to the distributions of $M(\Lambda_c^+\KS)$ and $M(\Lambda_c^+K^-)$.
In each case, the invariant-mass range between $2.91\gevcc$ to $3.15\gevcc$ is fit.
The PDF component that accounts for a possible resonance ($P_s$) is a double-Voigtian resonance shape with free parameters $\vec{a}$, which are the mass and natural width of the signal.
In the likelihood function, the means and widths of the resonance shapes are given Gaussian constraints (Eq.~(\ref{eq:Gauss})) based on the values measured with three-body decays.
For each of the five $\Lambda_c^+$ decay modes, the background distribution in $\Lambda_c^+\KS$ is modeled with a first-order polynomial ($P_b$).
Background distributions in $\Lambda_c^+K^-$ are modeled with functions proportional to Eq.~(\ref{eq:thresh}).

Invariant-mass resolutions and reconstruction efficiencies are calculated from MC samples.
Mass resolutions are about 2.5\mevcc.
Reconstruction efficiencies range from about 0.7\% to 5.0\% for signals that decay through $\Lambda_c^+\KS$, depending on the reconstructed $\Lambda_c^+$ decay mode, and from about 1.5\% to 12\% for signals that decay through $\Lambda_c^+K^-$.

\begin{figure}
\begin{minipage}[t]{0.5\textwidth}
\includegraphics[width=.95\textwidth]{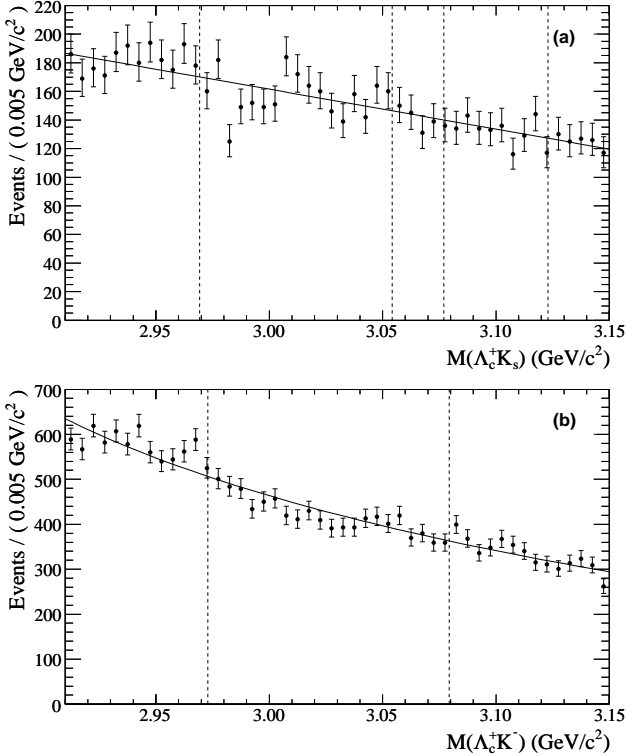}
\caption{Distributions of (a) $M(\Lambda_c^+\KS)$ and (b) $M(\Lambda_c^+K^-)$, for all five reconstructed $\Lambda_c^+$ decay modes combined.
The curves correspond to fits with no signal PDFs.
Signals were searched for around the invariant-mass values indicated by the dashed vertical lines.}
\label{fg:2KsK}
\end{minipage}
\end{figure}

Neither the $\Lambda_c^+\KS$ nor the $\Lambda_c^+K^-$ invariant-mass distributions exhibit evidence for statistically significant peaking structures corresponding to any of the charm-strange baryons found in the three-body final states.
Figure~\ref{fg:2KsK} shows background-only fits to the $\Lambda_c^+\KS$ and $\Lambda_c^+K^-$ invariant-mass distributions for all five reconstructed $\Lambda_c^+$ decay modes combined.
The same method described for three-body final states is used to determine 90\% confidence-level upper limits for the products of cross-section and branching fractions described by Eq.~(\ref{eq:cross}), where $Y$ is now $\Lambda_c^+\Kzb$ or $\Lambda_c^+K^-$. 
The upper limits on the products of cross-section and branching fractions are listed in Table~\ref{tb:24bodCross}.

\begin{table}
\caption{\label{tb:24bodCross} 90\% confidence-level upper limits on the products of cross-section and branching fractions (Eq.~(\ref{eq:cross})) for excited charm-strange baryons produced with $p^*>2.9\gevc$ and decaying to two-body and four-body final states.
NA indicates decays that are not kinematically allowed.
NO indicates that corresponding three-body decays are not observed.
The four-body final-state upper limits only use integrated yields up to about 30\mevcc above kinematic threshold. 
}
\begin{center}
\begin{tabular}{ccccccccc} \hline
$\Xi_c^*\rightarrow$    &\,& $\Lambda_c^+\Kzb$ &\,& $\Lambda_c^+K^-$  &\,& $\Lambda_c^+\Kzb\pi^-\pi^+$ &\,& $\Lambda_c^+K^-\pi^+\pi^-$ \\ \hline\hline
$\Xi_c(2980)$           &\,& $<12$\,fb                   &\,& $<10$\,fb         &\,& NA                         &\,& NA               \\
$\Xi_c(3055)$           &\,& $<\p9$\,fb                  &\,& NO                &\,& NA                         &\,& NA               \\
$\Xi_c(3077)$           &\,& $<2.9$\,fb                  &\,& $<1.2$\,fb        &\,& $<0.4$\,fb                 &\,& $<0.1$\,fb       \\
$\Xi_c(3123)$           &\,& $<2.7$\,fb                  &\,& NO                &\,& $<1.4$\,fb                 &\,& NO               \\ \hline\hline
\end{tabular}
\end{center}
\end{table}

\section{\boldmath Decays to $\Lambda_c^+\KS\pi^-\pi^+$ and $\Lambda_c^+K^-\pi^+\pi^-$}
\label{sec:4bod}

The studies of decays to the four-body final states $\Lambda_c^+\KS\pi^-\pi^+$ and $\Lambda_c^+K^-\pi^+\pi^-$ are based on fits to the one-dimensional distributions of $M(\Lambda_c^+\KS\pi^-\pi^+)$ and $M(\Lambda_c^+K^-\pi^+\pi^-)$.
The invariant-mass region from the minimum kinematic threshold ($\approx3063\mevcc$ for $\Lambda_c^+\KS\pi^-\pi^+$ and $\approx3059\mevcc$ for $\Lambda_c^+K^-\pi^+\pi^-$) to 250\mevcc above the threshold is fit in each case.
The PDF component that accounts for possible resonances is a double-Voigtian resonant function multiplied by a third-order polynomial that models the rapidly changing four-body phase space. 
In the likelihood function, the means and widths of the resonance shapes are given Gaussian constraints (Eq.~(\ref{eq:Gauss})) based on the values measured with three-body decays.
The rapidly rising four-body phase-space function and the proximity of any resonances to the kinematic thresholds lead to possibly significant contributions of $\Xi_c(3077)^0$ and $\Xi_c(3077)^+$ signals throughout each fitted 250\mevcc range.
In quoting upper limits for $\Xi_c(3077)^0$ and $\Xi_c(3077)^+$, we consider the integrated yield up to 3093\mevcc and 3089\mevcc, respectively ($\approx30\mevcc$ above threshold in each case).
Each range includes the peaking signal but not the higher mass range where efficiencies and possible effects from intermediate-resonant decays are unknown.
Background distributions are modeled by $(M-\alpha)(M-t)^2$, where $M$ is the mass variable, $t$ is the kinematic threshold, and $\alpha$ is a free parameter.

Invariant-mass resolutions and reconstruction efficiencies are again calculated from MC samples.
Mass resolutions are about 1.0\mevcc.
Reconstruction efficiencies range from about 0.3\% to 2.4\% for signals that decay through $\Lambda_c^+\KS\pi^-\pi^+$, depending on the reconstructed $\Lambda_c^+$ decay mode.
Reconstruction efficiencies range from about 0.8\% to 5.5\% for signals that decay through $\Lambda_c^+K^-\pi^+\pi^-$.

\begin{figure}
\begin{minipage}[t]{0.5\textwidth}
\includegraphics[width=.95\textwidth]{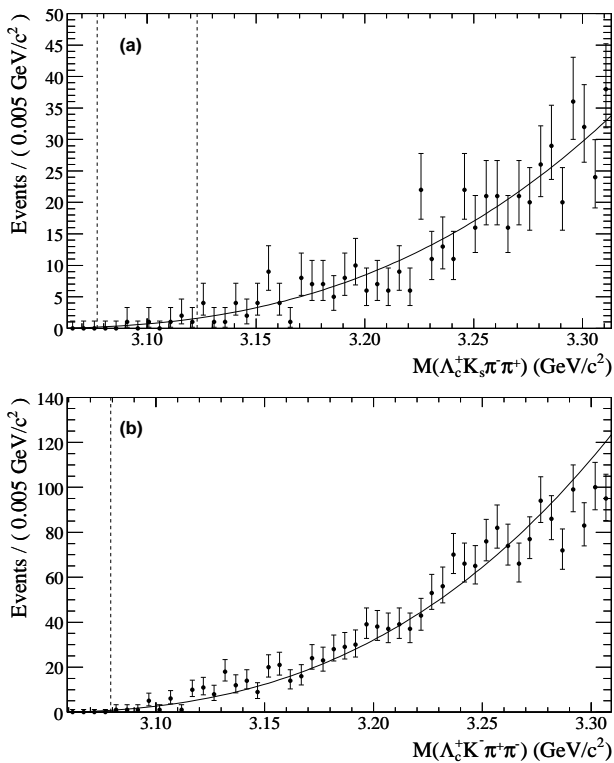}
\caption{Distributions of (a) $M(\Lambda_c^+\KS\pi^-\pi^+)$ and (b) $M(\Lambda_c^+K^-\pi^+\pi^-)$, for all five reconstructed $\Lambda_c^+$ decay modes combined.
The curves correspond to fits with no signal PDFs.
Signals were searched for around the invariant-mass values indicated by the dashed vertical lines.}
\label{fg:2KsKpipi}
\end{minipage}
\end{figure}

Neither the $\Lambda_c^+\KS\pi^-\pi^+$ nor the $\Lambda_c^+K^-\pi^-\pi^+$ invariant-mass distributions exhibit evidence for statistically significant peaking structures corresponding to any of the charm-strange baryons found in the three-body final states.
Figure~\ref{fg:2KsKpipi} shows background-only fits to these invariant-mass distributions for all five reconstructed $\Lambda_c^+$ decay modes combined.
The same method described for three-body final states is used to determine 90\% confidence-level upper limits for the products of cross-section and branching fractions described by Eq.~(\ref{eq:cross}), where $Y$ is now $\Lambda_c^+\Kzb\pi^-\pi^+$ or $\Lambda_c^+K^-\pi^+\pi^-$.
The upper limits on the products of cross-section and branching fractions are listed in Table~\ref{tb:24bodCross}.

\section{Summary and Conclusions}

Invariant-mass distributions for six different final states are studied for evidence of excited charm-strange baryons.
Four statistically significant signals are found for excited charm-strange baryons decaying to the final state $\Lambda_c^+K^-\pi^+$.
They are named $\Xi_c(2980)^+$ \mbox{($>9\,\sigma$)}, $\Xi_c(3055)^+$ ($6.4\,\sigma$), $\Xi_c(3077)^+$ \mbox{($>9\,\sigma$)}, and $\Xi_c(3123)^+$ ($3.8\,\sigma$),
and their masses, widths, and the products of cross-section and branching fractions are measured.
Intermediate-resonant $\Sigma_c(2455)^{++}K^-$ and nonresonant $\Lambda_c^+K^-\pi^+$ decays of the $\Xi_c(2980)^+$ are observed,
while the $\Xi_c(3077)^+$ is observed to decay mostly through the $\Sigma_c(2455)^{++}K^-$ and $\Sigma_c(2520)^{++}K^-$ intermediate-resonant states.
The $\Xi_c(3055)^+$ and $\Xi_c(3123)^+$ signals are observed only in $\Sigma_c(2455)^{++}K^-$ and $\Sigma_c(2520)^{++}K^-$ intermediate-resonant decays, respectively.
For the $\Lambda_c^+\KS\pi^-$ final state, the only statistically significant signal corresponds to the $\Xi_c(3077)^0$ baryon,
which is presumably the isospin partner to the $\Xi_c(3077)^+$;
its mass, width, and product of cross-section and branching fractions are measured.
Like the $\Xi_c(3077)^+$ state, the $\Xi_c(3077)^0$ is observed to decay mostly through the $\Sigma_c(2455)^{0}\KS$ and $\Sigma_c(2520)^{0}\KS$ intermediate-resonant states.
A $1.7\,\sigma$ enhancement in the final state $\Lambda_c^+\KS\pi^-$ is measured to have a mass and width consistent with its being the neutral isospin partner of the $\Xi_c(2980)^+$ baryon.
For the three-body final state $\Lambda_c^+\KS\pi^-$, 90\% confidence-level upper limits are determined for the products of cross-section and branching fractions of $\Xi_c(2980)^0$, $\Xi_c(3055)^0$, and $\Xi_c(3123)^0$.
No statistically significant signals are found in the two-body and four-body final states.
For states that are observed in three-body decays, we report 90\% confidence-level upper limits for the product of cross-section and branching fractions for kinematically allowed two-body and four-body decays.

The measured resonant and nonresonant decay rates and decay modes of these new excited charm-strange baryons might provide information on the internal quark dynamics.
In both the resonant and nonresonant strong decays, the strange quark is contained in a kaon, separate from the charmed baryon.
Previously known excited $\Xi_c$ baryons have been observed only in decays to a lower-mass $\Xi_c$ baryon plus a pion or photon.
Theoretically, there are several excited charm-strange baryon states with various spin, angular momentum, and radial excitation configurations~\cite{thspec,thlow,post1,post2,post3}.
These different theoretical states may offer explanations for differences in decay rates and decay modes, 
but the current theoretical and experimental information is not definitive enough to assign quantum numbers to any of the new excited charm-strange states.
Future experimental studies of these and other states and decay modes, and further theoretical work, will help to clarify the properties of these new charm-strange baryons. 

We are grateful for the extraordinary contributions of our \pep2\ colleagues in achieving the excellent luminosity and machine conditions that have made this work possible.
The success of this project also relies critically on the expertise and dedication of the computing organizations that support \babar.
The collaborating institutions wish to thank SLAC for its support and the kind hospitality extended to them.
This work is supported by
the US Department of Energy and National Science Foundation,
the Natural Sciences and Engineering Research Council (Canada),
the Commissariat \`a l'Energie Atomique and Institut National de Physique Nucl\'eaire et de Physique des Particules (France),
the Bundesministerium f\"ur Bildung und Forschung and Deutsche Forschungsgemeinschaft (Germany),
the Istituto Nazionale di Fisica Nucleare (Italy),
the Foundation for Fundamental Research on Matter (The Netherlands),
the Research Council of Norway,
the Ministry of Science and Technology of the Russian Federation,
Ministerio de Educaci\'on y Ciencia (Spain),
and the Science and Technology Facilities Council (United Kingdom).
Individuals have received support from
the Marie-Curie IEF program (European Union) and the A. P. Sloan Foundation.


\begin{thebibliography}{99}
\bibitem{conj} Throughout this paper, whenever a particle or decay mode is given, the charge conjugate is also implied.
\end{thebibliography}
\end{document}